\documentclass{article}
\usepackage{spconf,amsmath,graphicx}

\usepackage{adjustbox}
\usepackage{geometry}
\usepackage{amssymb}

\newgeometry{margin=2.5cm, left = 1.9cm, top = 2cm}
\title{Segmentation of Cortical Spreading Depression Wavefronts Through Local Similarity Metric}

\name{M. Filip Sluzewski$^{\dagger}$, Petr Tvrdik$^{\ddagger}$ and Scott T. Acton$^{\dagger}$, \textit{Fellow, IEEE}}
\address{$^{\dagger}$C.L. Brown Department of Electrical \& Computer Engineering, University of Virginia \\ $^\ddagger$ Center for Brain Immunology and Glia, Department of Neuroscience, University of Virginia}
\begin{document}
%
\maketitle
\begin{abstract}
In this paper, we present a novel region-based segmentation method for cortical spreading depressions in 2-photon microscopy images. Fluorescent microscopy has become an important tool in neuroscience, but segmentation approaches are challenged by the opaque properties and structures of brain tissue. These challenges are made more extreme when segmenting events such as cortical spreading depressions, where low signal-to-noise ratios and intensity inhomogeneity dominate images. The method we propose uses a local intensity similarity measure that takes advantage of normalized Euclidean and geodesic distance maps of the image. This method provides a smooth segmentation boundary which is robust to the noise and inhomogeneity within cortical spreading depression images. Experimental results yielded a DICE index of 0.9859, an increase of 6\% over the current state-of-the-art, and a reduction of root mean square error by 79.9\%.
\end{abstract}

\section{Introduction}
\label{sec:intro}

\textit{In vivo} imaging of the central nervous system has become an increasingly prevalent approach to neuroscience research. Advances in transgenic animal models and two-photon microscopy have allowed researchers to directly observe the development and behavior of many cells and structures within the brain which hitherto have only been approximated through \textit{in vitro} studies and indirect observation \cite{friston1998imaging,tvrdik2017vivo,svoboda2006principles}. However, due to the complexities of the tissues being imaged, this technique has several challenges which affect image quality. The primary challenge is that tissue is adept at scattering light, which results in noise and decreasing fluorescence intensity as the imaging depth increases \cite{oheim2001two}. This problem of noise and clarity is compounded by the fluorescence of tissue outside the focal plane, as well as partial or total occlusion of objects in the image by structures that lie above the focal plane, such as blood vessels. Because of these difficulties, the development of robust image segmentation strategies, both for general and specific applications, is critical to successful analysis.
\par
Traditional approaches to segmentation of biological images such as active contours \cite{ray2002active,mansouri2004constraining}, level set methods \cite{acton2009biomedical}, and watershed techniques \cite{wright1997watershed} fail due to poor contrast and inhomogeneity of the observed intensity. In this paper, we propose a method for segmenting the wavefront of cortical spreading depressions (CSDs) in two-photon calcium images of mouse brains. This method is inspired by the region-based, noise-robust method proposed by Nui \textit{et al}. \cite{niu2017robust} and converts a level set method to a threshold-based method that is essentially based on fast marching \cite{sethian1999level}. The primary motivation of the proposed method is to conserve the shape of the segmentation boundary between iterations, preventing the final segmentation from being overly sensitive to the discontinuities in the CSD wavefront that occur due to noise and occlusion. Furthermore, the proposed method seeks not to segment an object in an image from the background in the traditional sense, but rather to separate the image functionally into two regions: one where the neurons have depolarized in response to the CSD and one where they have yet to depolarize.
\section{Background}
\label{sec:background}

\subsection{Cortical Spreading Depressions}

\begin{figure}[b!]
	\centering
	\renewcommand{\tabcolsep}{0.05cm}
	\setlength{\belowcaptionskip}{-5pt}
	\begin{adjustbox}{width=\linewidth}
	{
			\begin{tabular}{cc}
				\includegraphics[width=.18\linewidth, height = 0.13\linewidth]{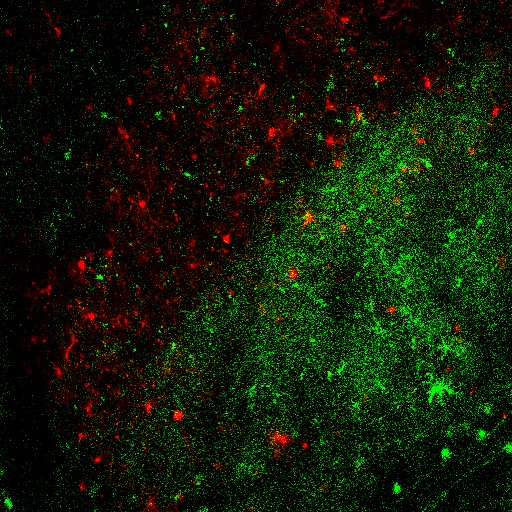}&
				\includegraphics[width=.18\linewidth, height = 0.13\linewidth]{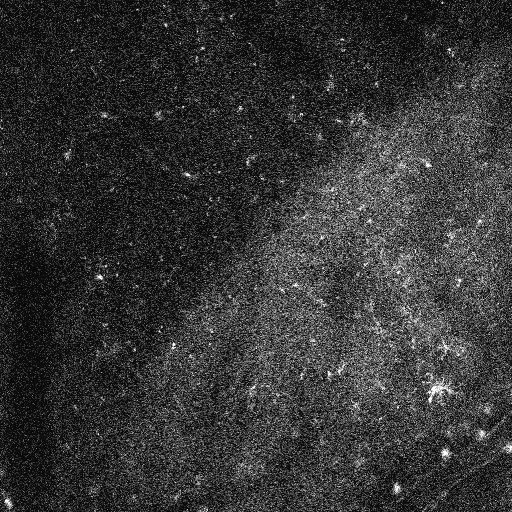}				
			\end{tabular}
		}	
	\end{adjustbox}
	\vspace{-0.3cm}
	\caption{(\textit{left}) Multi-photon confocal microscopy image of a cortical spreading depression (CSD). Microglia (red) respond to calcium signalling from the rapid depolarization of neurons (green) \cite{tvrdik2017vivo}. (\textit{right}) Isolated green-channel image of CSD. Brighter regions indicate areas which have already undergone depolarization.}
	
	\label{fig:example}
\end{figure}

CSDs are a pathological neurological phenomena associated with traumatic brain injury, stroke, and migraines. They are characterized by a slowly-propagating (1.7-9.2mm/min) wave of near-complete neuron depolarization, followed by a period of suppressed neural activity. In traumatic brain injuries and stroke, CSDs occur repeatedly with increasing intensity over time, damaging healthy neurons and exacerbating the injury \cite{dreier2017recording,fabricius2005cortical}. Because of this, research into the causes of CSDs, the neurological responses they induce, and methods of inhibiting their generation and propagation are important to improving patient outcome.
\par
While CSDs and their role in brain injury have been investigated in previous studies, these primarily relied on the use of electrodes for \textit{in vivo} recording of the generation and propagation of CSDs \cite{dreier2017recording}. In recent years, researchers have begun to use two-photon microscopy to image CSDs \cite{tvrdik2017vivo}. These imaging studies are particularly useful in investigating the suspected role of non-neuronal cells, such as microglia, in the propagation of CSDs and the resulting cellular response \cite{tvrdik2017vivo,mizuma2018microglial,szalay2016microglia}. In these studies, the propagation of the CSD is detected through calcium indicators (such as GCaMP) which fluoresce in response to increased intracellular calcium \cite{stosiek2003vivo}. In this paradigm, the CSD appears as a region of increased fluorescence as neuronal dendrites depolarize in the focal plane, which rapidly propagates as a wavefront across the image (see Fig. 1). 
\par 
Accurately segmenting the boundary of the CSD can provide valuable information about which regions have undergone depolarization in a given image, as well as measuring the direction and speed of the CSD. However, these imaging sequences are often characterized by a low signal-to-noise ratio (SNR), intensity inhomogeneity, and discontinuities in the visible wavefront as a result of light scattering and occlusion. These obstacles make delineating the wavefront difficult for many segmentation methods which rely on edge detection or global information. This is because the low SNR can make it challenging to accurately calculate edges within the image, while the intensity inhomogeneity and discontinuities make it challenging to calculate a global statistic that properly distinguishes the depolarized regions of the image from the rest of the image. This results in segmentations where the contour boundary fails to advance towards the CSD wavefront, or ones which "sink into" the discontinuous or inhomogeneous regions of the wavefront, resulting in erroneous segmentation. Because of this, we seek to implement a region-based method which is robust to such noise and inhomogeneity, thus resulting in a segmentation that accurately converges to the wavefront boundary in a smooth and continuous manner. 

\subsection{Local Similarity}
The inspiration behind the method proposed in this paper is the local similarity factor (LSF). First introduced by Nui \textit{et al.} \cite{niu2017robust}, the LSF provides a distance-weighted, region-based measurement of the similarity between the intensities of pixels within a region and the region's mean intensity. For a given pixel, $x$, within an image, $I$, the LSF is defined as:
\begin{equation} 
LSF(x,lc)=\int_{y\in N_{x}\neq x}\frac{\left | I(y)-lc \right |^{2}}{d(x,y)}dy 
\end{equation}
where $N_{x}$ is a square-shaped window defining the local region, $d(x,y)$ is the Euclidean distance between pixels $x$ and $y$, and $lc$ is the local mean intensity. The LSF metric has two major advantages that are valuable for the problem addressed in this paper. First, the method does not require preprocessing prior to segmentation, in the form of noise reduction or contrast enhancement, for the metric to be effective. Second, the LSF-based model the authors present is robust to high levels of noise as well as intensity inhomogeneity, making LSF an appealing metric to apply to the CSD images this paper focuses on \cite{niu2017robust}. 
\par
Two different LSF values are computed per pixel for each iteration of the algorithm. $LSF_{1}$ compares pixels that are inside the current segmentation boundary while $LSF_{2}$ compares pixels that are outside the current segmentation boundary. As a result, two local mean intensities, $lc_{1}$ and $lc_{2}$, are calculated when evaluated over a level set $\phi(\cdot)$:
\begin{equation} 
lc_{1}(x)=\frac{\int_{\Omega}M(x,y)I(y)H_{\varepsilon}(\phi(y))dy}{\int_{\Omega}M(x,y)H_{\varepsilon}(\phi(y))dy} 
\end{equation}
and
\begin{equation} 
lc_{2}(x)=\frac{\int_{\Omega}M(x,y)I(y)(1-H_{\varepsilon}(\phi(y)))dy}{\int_{\Omega}M(x,y)(1-H_{\varepsilon}(\phi(y)))dy}. 
\end{equation}
$M(x,y)$ is a mask of the local region defined as
\begin{equation}
M(x,y)\left\{\begin{matrix} 1 & d(x,y)<r\\ 0 & else\end{matrix}\right.
\end{equation}
and $H_{\varepsilon}(\cdot)$ is the regularized Heaviside function. 
\par
By combining equations (1-4), Nui \textit{et al.} produce the following energy functional, called the Region-based model via Local Similarity Factor (RLSF):
\begin{equation}
\begin{split}
&E^{RLSF}(x,\phi(x))=\\
& = \lambda_{1}\int_{\Omega}LSF_{1}(x,lc_{1}(x))H_{\epsilon}(\phi(x)) dx \\
& + \lambda_{2}\int_{\Omega}LSF_{2}(x,lc_{2}(x))(1-H_{\epsilon}(\phi(x)))dx \\
& + \mu\int\delta_{\epsilon}(\phi(x)) | \triangledown \phi(x) |  dx
\end{split}
\end{equation}
where $\delta_{\epsilon}(\cdot)$ is the regularized Dirac delta function and $\lambda_{1}$, $\lambda_{2}$, and $\mu$ are weighting terms. The last integral of the energy functional serves as a smoothing parameter, with larger values of $\mu$ resulting in a smoother contour.

\section{Method: Local Similarity Metric}

\label{sec:method}
\subsection{Distance map-based segmentation with LSM}
\label{ssec:locSimThresh}
\par
In order to compensate for the high level of noise, intensity inhomogeneity, and wavefront occlusion present in confocal images of CSDs, we propose an modified version of the RLSF model which recontextualizes it from an active contour to a threshold-based approach, which we call the Local Similarity Metric (LSM).  
\par
In the LSM model, the boundary of a given threshold is defined according to a distance map representation of the image rather than the pixel intensities of the image. This normalized distance map, $D_{I}(x)$, is defined as:
\begin{equation}
    D_{I}(x) = (2H_{\epsilon}(\phi_{0}(x))-1) \frac{d_{0}(x)g_{0}(x)}{max_{y\in \Omega}\left \{ d_{0}(y)g_{0}(y) \right \}}
\end{equation}
where $d_{0}$ and $g_{0}$ are normalized Euclidean and geodesic distance maps of the image, respectively, and $\phi_{0}(\cdot)$ is the user-defined initial contour. The initial contour's boundary pixels, excluding those on the edges of the image, serve as the "zero-points" of both distance maps, meaning that the value of a pixel $x$ is the shortest distance from it to one of the boundary pixels. In order to encourage a smooth curve shape over various threshold values, a median filter is applied to the map. 
\par 
Using (6), we can define a pseudo-level set of the image, with respect to a given threshold $T$, as such:
\begin{equation}
    \phi(x,T)=\left\{\begin{matrix}
1 & D_{I}(x)\leq T\\ 
-1 & else.
\end{matrix}\right.
\end{equation}
Using this psuedo-level set, we define the LSM energy functional as follows:
\begin{equation}
\begin{split}
&E^{LSM}(T)=\\
&\lambda_{1}\int_{\Omega}LSF_{1}(x,lc_{1}(x))H_{\epsilon}(\phi(x,T))dx\\
&+ \lambda_{2}\int_{\Omega}LSF_{2}(x,lc_{2}(x))(1-H_{\epsilon}(\phi(x,T)))dx.
\end{split}
\end{equation}
Note that the smoothing term has been dropped from the original RLSF model (5), as the contour shape is strictly defined by $D_{I}$ and the choice of $T$. A smooth contour is enforced through the use of Euclidean distance in generating $D_{I}$ and the median filter applied to it. Furthermore, both $lc_{1}$ and $lc_{2}$, in (2-3), also use $\phi(x,T)$ instead of $\phi(x)$. For the sake of brevity, we will not explicitly redefine them here.

\subsection{Threshold optimization through gradient descent}
\label{ssec:threshold}

Once the initial contour has been defined and the appropriate distance map has been generated, the optimal threshold value for the image is calculated through the use of a gradient descent algorithm with a fast marching method-like implementation. \par
The partial differential equation used for the gradient descent algorithm is as follows:
\begin{equation}
    \frac{\partial T}{\partial t}=\lambda_{1}\int_{N_{T}}LSF(x,lc_{1})dx-\lambda_{2}\int_{N_{T}}LSF(x,lc_{2})dx,
\end{equation}
where $N_{T}$ is a subset of the image containing pixels that are within a specified distance from the current threshold boundary. However, the curved geometry of the CSD wavefront often leads to an imbalance within $N_{T}$ between the number of pixels outside versus inside the current threshold, which biases one side over the other. We equalize this imbalance by removing pixels on the overrepresented side from $N_{T}$.
\par
In order to minimize (8), the following gradient descent formulation is utilized:
\begin{equation}
    T_{n+1}= T_{n}+\Delta t\Delta T_{n}
\end{equation}
where $\Delta T_{n}$ is a numerical approximation of (9). $\Delta t$ is the step-size, defined as:
\begin{equation}
    \Delta t=\frac{1}{Nmax_{ x\in N_{T} }( LSF_{1}(x,lc_{1})-LSF_{2}(x,lc_{2}))}
\end{equation}
where $N$ is the number of pixels within $N_{T}$.

\section{Experimental Results}
\label{sec:results}

\begin{figure*}[ht!]
	\centering
	\renewcommand{\tabcolsep}{0.05cm}
	\setlength{\belowcaptionskip}{-10pt}
	\begin{adjustbox}{width=.85\textwidth}
	{
			\begin{tabular}{ccccc}
				\includegraphics[width=.12\linewidth, height = 0.1\linewidth, scale=0.1]{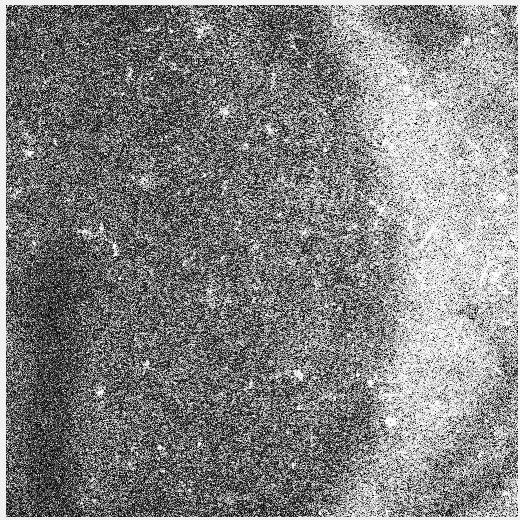} &
				\includegraphics[width=.12\linewidth, height = 0.1\linewidth, scale=0.1]{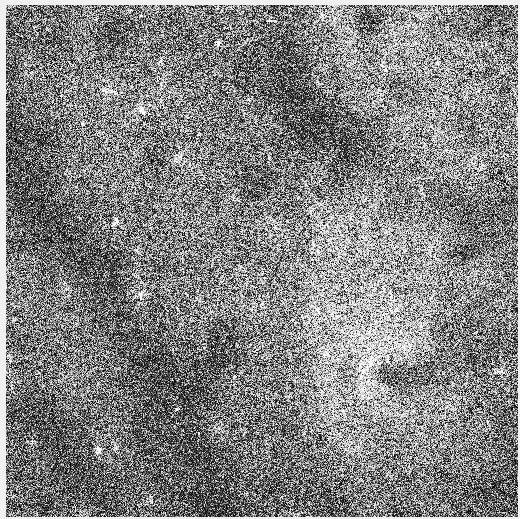} &
				\includegraphics[width=.12\linewidth, height = 0.1\linewidth, scale=0.1]{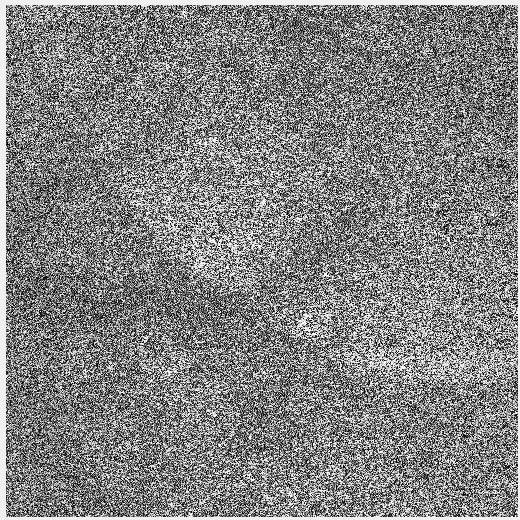} &
				\includegraphics[width=.12\linewidth, height = 0.1\linewidth, scale=0.1]{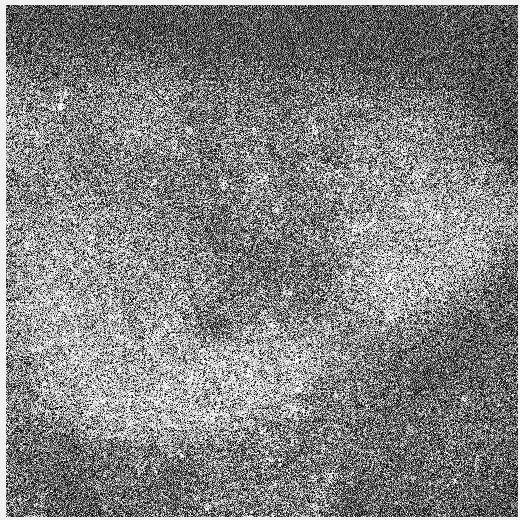} &
				\includegraphics[width=.12\linewidth, height = 0.1\linewidth, scale=0.1]{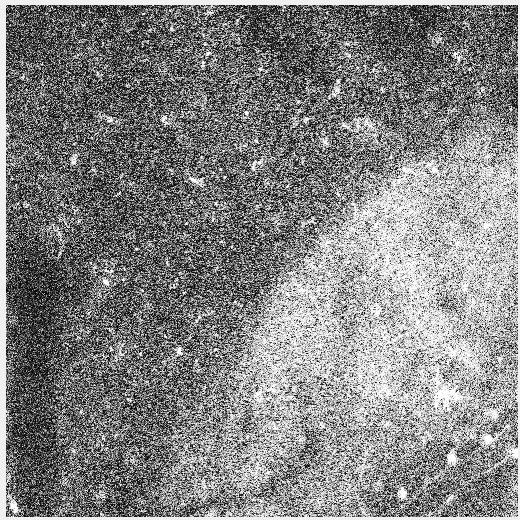} 
					\\
					
				\includegraphics[width=.12\linewidth, height = 0.1\linewidth, scale=0.1]{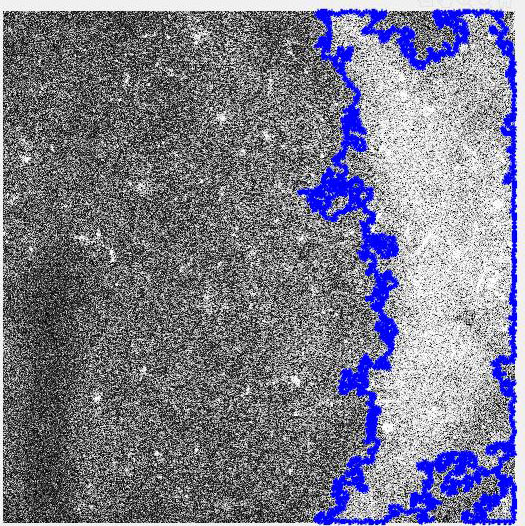} &
				\includegraphics[width=.12\linewidth, height = 0.1\linewidth, scale=0.1]{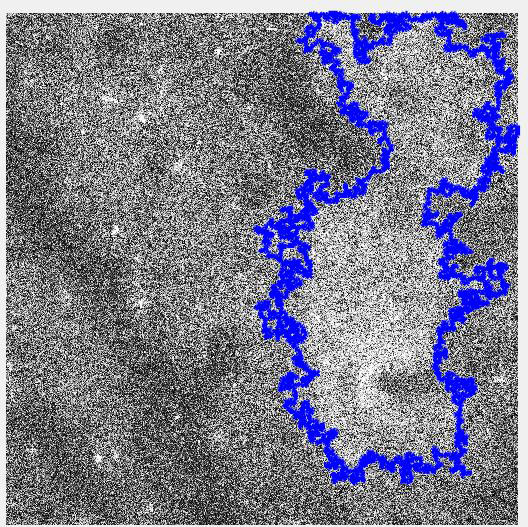} &
				\includegraphics[width=.12\linewidth, height = 0.1\linewidth, scale=0.1]{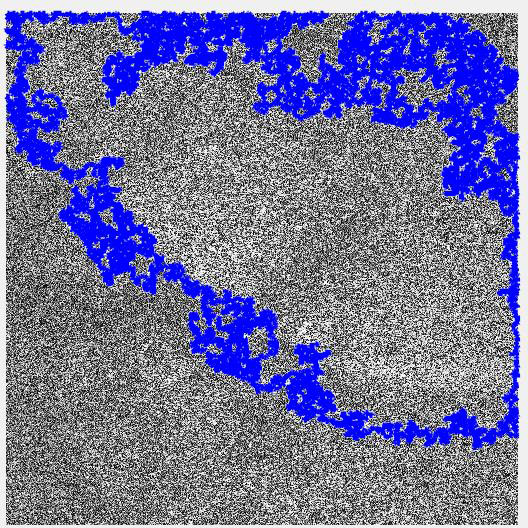} &
				\includegraphics[width=.12\linewidth, height = 0.1\linewidth, scale=0.1]{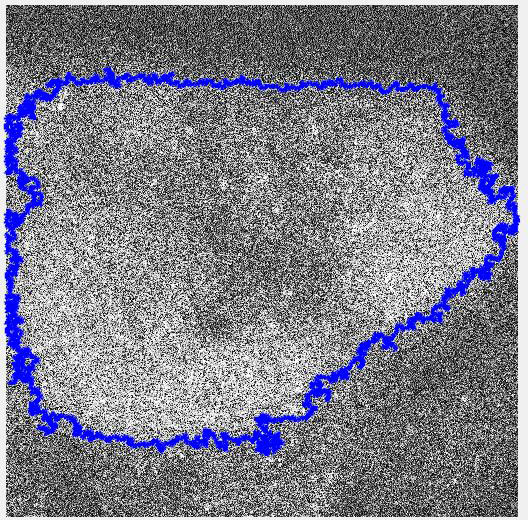} &
				\includegraphics[width=.12\linewidth, height = 0.1\linewidth, scale=0.1]{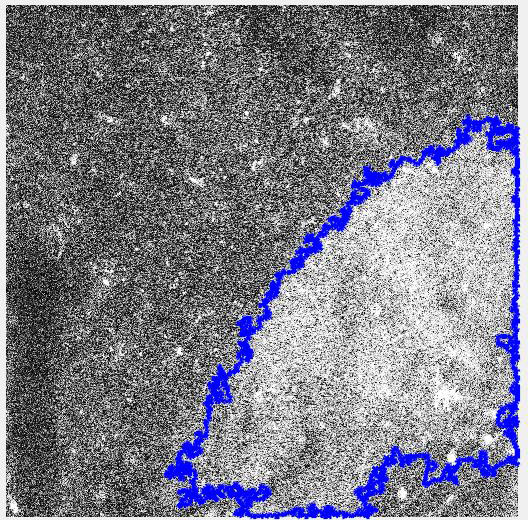} 
				\\
					
				\includegraphics[width=.12\linewidth, height = 0.1\linewidth, scale=0.1]{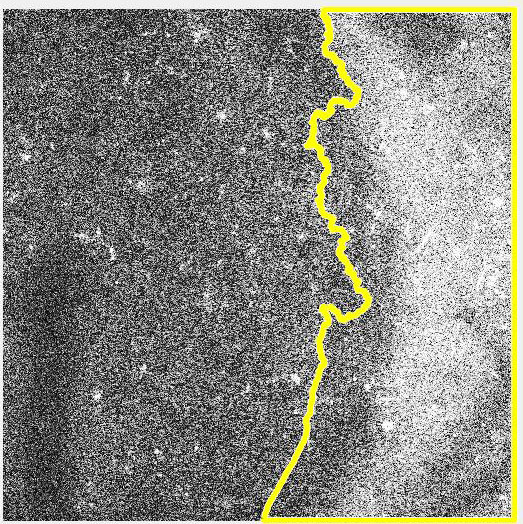} &
				\includegraphics[width=.12\linewidth, height = 0.1\linewidth, scale=0.1]{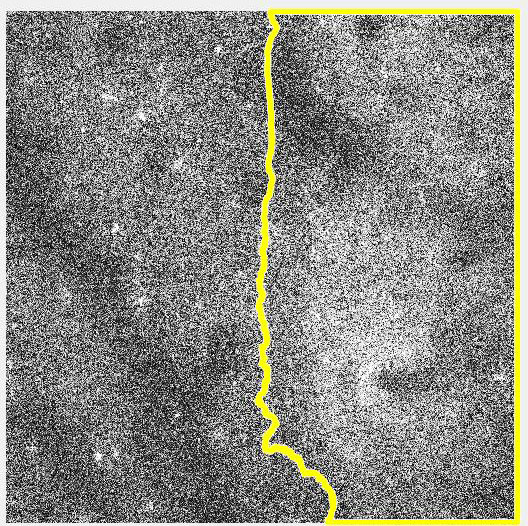} &
				\includegraphics[width=.12\linewidth, height = 0.1\linewidth, scale=0.1]{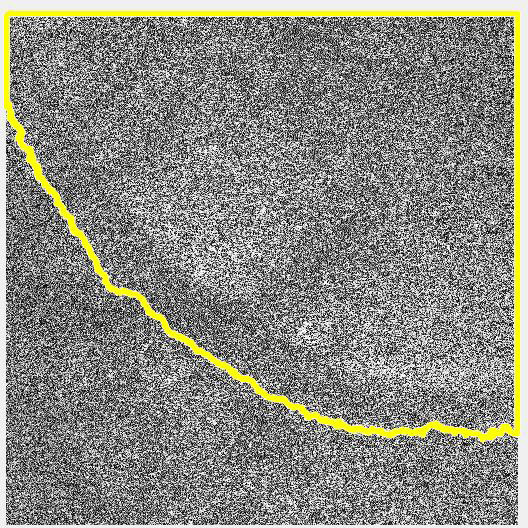} &
				\includegraphics[width=.12\linewidth, height = 0.1\linewidth, scale=0.1]{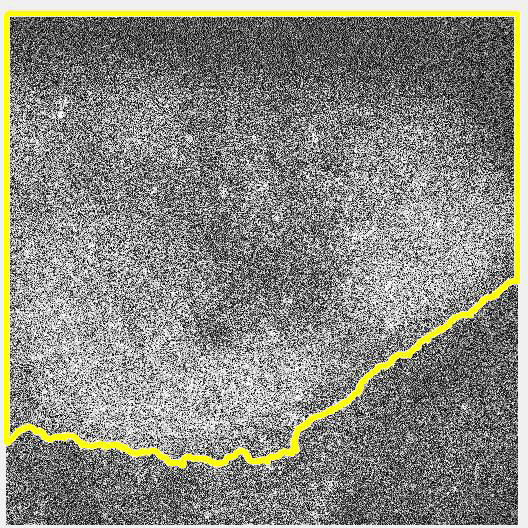} &
				\includegraphics[width=.12\linewidth, height = 0.1\linewidth, scale=0.1]{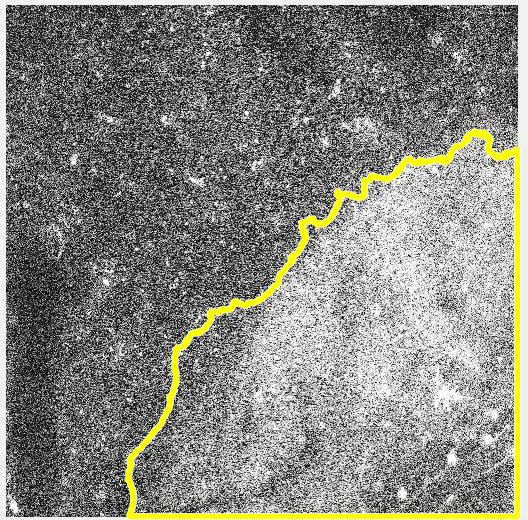} 
				\\
				
				\includegraphics[width=.12\linewidth, height = 0.1\linewidth, scale=0.1]{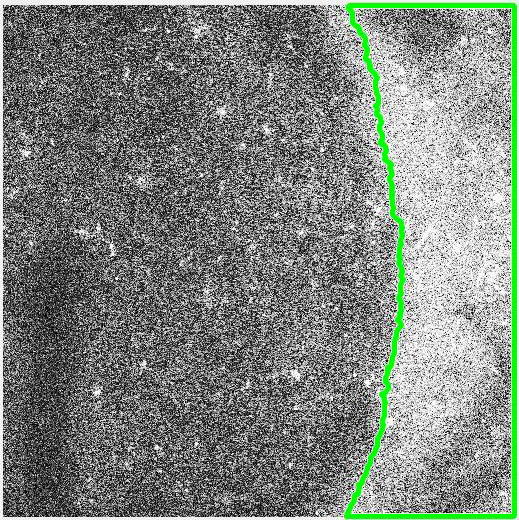} &
				\includegraphics[width=.12\linewidth, height = 0.1\linewidth, scale=0.1]{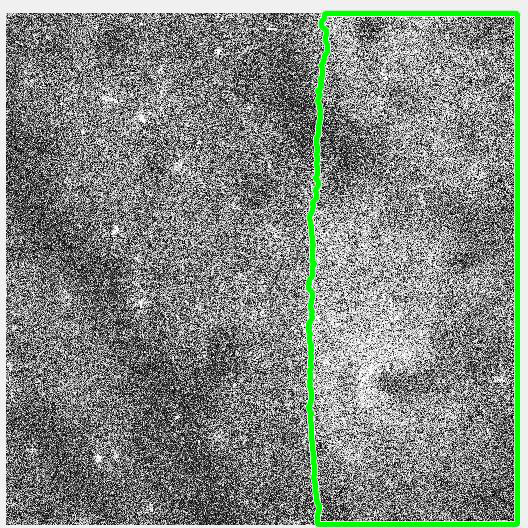} &
				\includegraphics[width=.12\linewidth, height = 0.1\linewidth, scale=0.1]{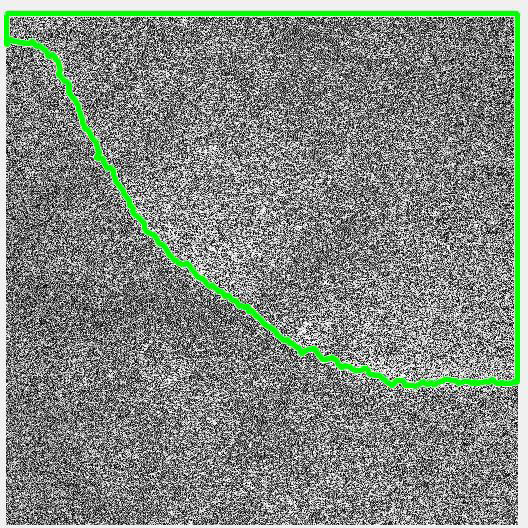} &
				\includegraphics[width=.12\linewidth, height = 0.1\linewidth, scale=0.1]{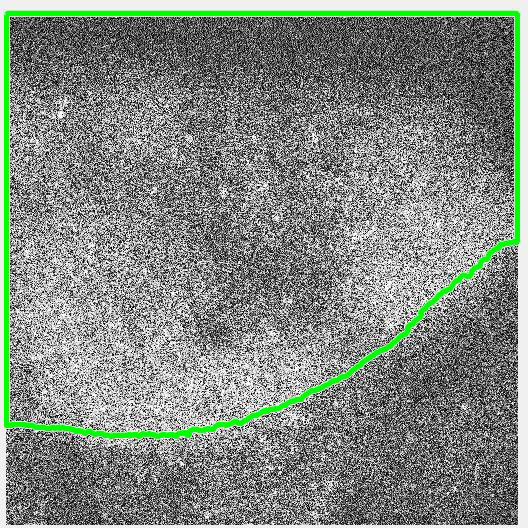} &
				\includegraphics[width=.12\linewidth, height = 0.1\linewidth, scale=0.1]{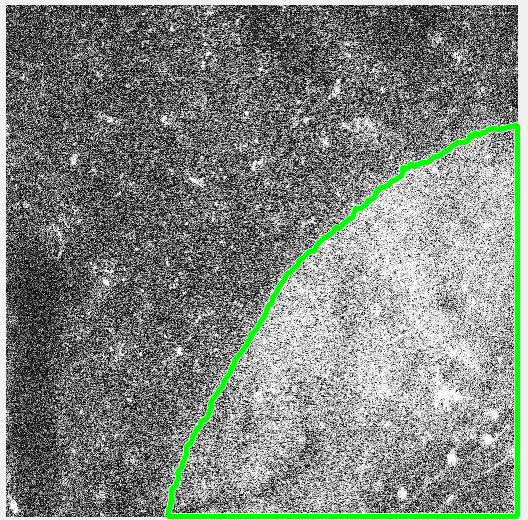}
		
			\end{tabular}
		}	
	\end{adjustbox}
	\caption{Segmentation results of CSD images. \textit{(First row)} Raw calcium images of CSD. \textit{(Second row)} Chan-Vese \cite{chan2000active}. \textit{(Third row)} Mukherjee \textit{et al}. \cite{mukherjee2015region}. \textit{(Fourth row)} LSM. For display purposes, images have undergone contrast enhancement, which was not performed prior to segmentation.}
	
	\vspace{-0.1cm}
	\label{fig:results}
\end{figure*}

\subsection{Dataset and parameter selection}
\label{ssec:dataAndParam}

In order to evaluate the performance of our proposed method, we compiled a dataset of twenty 512x512 pixel images of CSDs taken from two-photon confocal calcium imaging sequences of mouse brains. CSDs were induced in the mice either through surgical occlusion of the middle cerebral artery (MCA) or a 1 M injection of KCl directly into the cortex. Images from these sequences were selected such that the dataset contained a variety of possible noise and contrast levels, as well as various shapes that the CSD wavefront can assume. While we would have liked to have had a larger dataset for this analysis, the novelty of this imaging technique and the overall cost of such experiments limits the quantity and quality of sequences at our disposal.
\par
As stated before, our algorithm requires several parameters to be specified by the user prior to segmentation. Given the resolution of our imaging instrument, for the local window, $N_{x}$, we selected a size of 17x17 pixels, while the regional mask, $M$, had a radius of 13 pixels. For the region used to calculate the gradient descent, $N_{T}$, we considered pixels within 7 pixels of the current boundary. Finally, the weighting terms were set as $\lambda_{1}=\lambda_{2}=1$. The geodesic distance map used in creating $D_{I}$ was generated using the MATLAB (MathWorks, CA) command \texttt{immsegfmm}.
\subsection{Performance Evaluation}
\label{ssec:performance}

To evaluate the performance of the LSM method, we compared the results of our approach with that of Chan-Vese \cite{chan2000active} and Mukherjee \textit{et al}. \cite{mukherjee2015region}. The implementation of Mukherjee \textit{et al}. method that we used in this evaluation utilizes a fast-marching method approach which prevents contour edges that lie on the edges of an image from evolving. Chan-Vese segementation was performed with MATLAB's built in implementation, \texttt{activecontour}. Both Mukherjee \textit{et al}. and Chan-Vese were executed for 1000 iterations, while LSM ran for only 50 iterations.
\par 
Performance was measured using two different metrics. The first was the DICE index, which measures the overlap between the segmentation result and the ground truth \cite{bernard2009variational}. This is calculated as $DICE(R_{1},R_{2})=\frac{2Area(R_{1}\cap R_{2})}{Area(R_{1})+Area(R_{2})}$ where $R_{1}$ is the ground truth and $R_{2}$ is the segmentation. A DICE index closer to 1 indicates superior performance. However, due to the large area of the image being segmented, a poor segmentation can still result in a large DICE index. Therefore, we used the root mean square error (RMSE) as a second performance metric. Here, the error being measured is the Euclidean distance between a given point on the segmentation boundary and the nearest point on the ground truth boundary. For the initialization, an initial contour was hand-drawn such that the only edge of the contour which did not lie on the image boundary was a rough approximation of the CSD wavefront, displaced approximately 40 pixels away from the wavefront boundary in the image. These initial segments had a mean DICE index of .9213 and a mean RMSE of 23.00 pixels.
\par
Over the entirety of the dataset, our proposed method demonstrated superior performance compared to the other methods examined (Fig. 2). The LSM yeilded a mean DICE index of .9859, compared to that of .9308 and .8136 for Mukherjee \textit{et al}. and Chan-Vese, respectively. Likewise, LSM had a mean RMSE of 4.52 pixels, compared to 22.57 pixels and 30.19 pixels for Mukherjee \textit{et al}. and Chan-Vese, respectively. Failures of Mukherjee \textit{et al}.'s method were largely characterized by the contour's inability to evolve towards the wavefront. By contrast, intensity inhomogeneity was the primary cause of failure for Chan-Vese. The active contour would converge around the brightest portions of the depolarized region, often excluding regions where noise and occlusion obscured the increased fluorescence. 

\section{Conclusion}
\label{sec:conlcusion}
In this paper, a novel approach to segmenting wavefront boundaries in two-photon calcium images of CSDs is presented. Qualitative analysis of the performance for LSM indicates improvements in comparison to the state of the art. In addition to accurate segmentation results, the LSM method has a fast rate of convergence, requiring less than 50 iterations. The most significant downside of this method is its dependence on the user's \textit{a priori} knowledge of the wavefront's shape, which limits throughput on larger datasets. Inclusion of the LSM into models which can incorporate \textit{a priori} shape information in an adaptable manner is an appealing solution for future work.


\bibliographystyle{IEEEtran}
\bibliography{refs}

\end{document}